\documentclass[11pt,a4paper]{article} 
\usepackage{jheppubori} 
\usepackage{hyperref} 
\usepackage{diagbox}
\usepackage{comment}
\usepackage{bigints}
\usepackage{ulem}
\usepackage{amsmath}
\usepackage{graphicx}
\usepackage{subfigure}
\usepackage{amssymb}
\usepackage{multirow}
\usepackage{cancel}
\usepackage{color}
\usepackage{mathtools}
\usepackage{listings}
\usepackage{xcolor}
\usepackage{float}
\usepackage{slashed}
\usepackage{bm}
\usepackage{wrapfig}
\usepackage{lineno}
\usepackage{marginnote} 
\usepackage[toc,title,page]{appendix}
\usepackage[colorinlistoftodos]{todonotes}

\newcommand{\tata}{\tau^+ \tau^-}
\newcommand{\mumu}{\mu^+ \mu^-}

\title{\boldmath The effects of a scalar singlet Leptoquark at the $Z$ factory}

\author[a]{Dazhuang He,}
\author[b, 1]{Yu Zhang,~\note{Corresponding author.}}
\author[c, 2]{and Hao Sun~\note{Corresponding author.}}

\affiliation[a]{College of Physics and Electronic Engineering, Heze University,\\ No.2269 University Road, Heze, Shandong, 274015, P.R.China}
\affiliation[b]{School of Physics, Hefei University of Technology, No.420 Feicui Road,\\ Hefei, Anhui, 230601, P.R.China}
\affiliation[c]{Institute of Theoretical Physics, School of Physics, Dalian University of Technology, \\ No.2 Linggong Road, Dalian, Liaoning, 116024, P.R.China}

\emailAdd{dzhe1998@gmail.com}
\emailAdd{dayu@hfut.edu.cn}
\emailAdd{haosun@dlut.edu.cn}

\abstract{
We evaluate the observability of the effects of a scalar singlet leptoquark (LQ) in $\mu$ and $\tau$-pair productions at the $Z$ factory.
In the scenario addressing the charged-current anomalies, the LQ contributions to $\mu$-pair final state are negligible.
In contrast, a sizable contribution arises in the $\tau$-pair production, which is identical in both $Z$ decay and $e^+e^-$ collider at $Z$ pole.
These effects are mainly sensitive to left-handed interaction, showing a maximum deviation of about $-0.7\%$ for both 1\,TeV and 2\,TeV LQ. 
The suppression of new physics effects from the heavy LQ can be compensated by the enlarged couplings parameter space.
For the $\tau$-pair production channel, we further specify the coupling constraints corresponding to the expected measurement precision at the future $Z$ factory. 
Moreover, we provide an analytic function in terms of the LQ mass and couplings to quantify the LQ effects.
The differential distributions in the collision process indicate that the LQ effects remain stable throughout the kinematic region. 
Meanwhile, the measurement sensitivity of the $\tau$-pair final state at the future $Z$ factory is expected to impose further constraints on the LQ theory.
}

\begin{document}
\maketitle
\flushbottom
%%%%%%%%%%%%%%%%%%%%%%%%%%%%%%%%%%%%%%%%%%%%%%%%%%%%%%%%%%%%%%%%%%%%%%%%%%%%%%%%%%%%%%%%%%%%%%%%%%%%%%%%%%%%%%%%%%%%%%%%%%%%%%%%%%%%%%
%%%%%%%%%%%%%%%%%%%%%%%%%%%%%%%%%%%%%%%%%%%%%%%%%%%%%%%%%%%%%%%%%%%%%%%%%%%%%%%%%%%%%%%%%%%%%%%%%%%%%%%%%%%%%%%%%%%%%%%%%%%%%%%%%%%%%%
\section{Introduction}
\label{sec:intro}

The Standard Model (SM) provides a fundamental theoretical framework for describing the interactions between elementary particles, especially following the discovery of the Higgs boson at the Large Hadron Collider (LHC) in 2012~\cite{ATLAS:2012yve,CMS:2012qbp}. 
However, several experimental observations indicate that the SM is insufficient to fully account for certain phenomena. 
It should be regarded as an effective theory of higher energy scales ($\sim$ TeV).  
As a consequence, the LHC has shifted its focus toward searching for deviations from SM, which may indicate the new physics (NP). 
Although, the direct searches at the LHC have not yet provided conclusive evidence for NP, 
the experimental observations in B-meson decays have revealed significant deviations from SM expectations, such as the phenomena of lepton flavor universality violation (LFUV). 
LFUV was first reported in B-meson semileptonic decays by the BaBar collaboration~\cite{BaBar:2012obs,BaBar:2013mob}, and has subsequently been updated and confirmed by the Belle and LHCb collaborations~\cite{LHCb:2017avl,LHCb:2017smo,LHCb:2017rln,Belle:2019rba,LHCb:2019hip,Belle:2019oag,BELLE:2019xld,LHCb:2021trn,Belle-II:2024ami,LHCb:2023uiv,LHCb:2023zxo}. 
For example, the observables $R(D^{(*)})$ are introduced to describe LFUV in B-meson semileptonic decays with $D^{(*)}$ final states
\begin{equation}
\begin{aligned}
R(D^{(*)})=\frac{\Gamma (B\to D^{(*)} \tau \nu)}{\Gamma (B\to D^{(*)} l \nu)}, \quad l = e,\mu.   
\end{aligned}
\end{equation}
The results of \( R(D^{(*)}) \) are presented in Table.~\ref{tab::Rds}, where the experimental measurements exhibit significant deviations from SM predictions.

An alternative approach to addressing LFUV is to introduce new composite particles, known as leptoquarks~(LQs)~\cite{Dorsner:2013tla,Sakaki:2013bfa,Bauer:2015knc,Barbieri:2015yvd}.
LQs carry color charge, as well as both lepton and baryon quantum numbers.  
Consequently, LQs can couple directly to a quark and a lepton via Yukawa-type interactions at tree level. 
For instance, as shown in FIG.~\ref{fig::LQ}, the introduction of a scalar singlet LQ ($S_1$) can contribute to the charged-current (CC) transition \( b \to c \tau \nu \), thereby generating additional effects beyond those predicted by the SM. 
The theory of LQs has been extensively investigated in the previous works~\cite{Mandal:2018kau,Crivellin:2021egp,Crivellin:2021bkd,Borschensky:2022xsa,Buttazzo:2017ixm,Marzocca:2018wcf,Gherardi:2020qhc,Bhaskar:2023ftn,Vignaroli:2018lpq,Greljo:2017vvb,He:2025arp,Alvarez:2018jfb,Haisch:2022afh,Haisch:2022lkt,Buonocore:2020erb,Raj:2016aky,Bansal:2018eha}. 
In particular, by matching the LQs theory to the Standard Model Effective Field Theory (SMEFT), the contributions of scalar LQs to low-energy observables have been calculated at the one-loop level in Ref.~\cite{Gherardi:2020qhc}. 
A systematic description and implementation of the complete scalar LQs Lagrangian with {\tt FeynRules} package~\cite{Christensen:2008py} has been presented in Ref.~\cite{Crivellin:2021ejk}. 
In the scenarios addressing $(g-2)_{\mu/\tau}$, the LQ effects in $Z$ decays and electro-positron colliders are investigated in Refs.~\cite{Crivellin:2020mjs,ColuccioLeskow:2016dox,Crivellin:2021spu,Arnan:2019olv}.
At LHC, the NLO QCD corrections to scalar LQ pair production have been calculated in Refs.~\cite{Kramer:1997hh,Kramer:2004df} and further matched with the parton shower in Refs.~\cite{Ghosh:2023ocz,Borschensky:2021jyk,Borschensky:2021hbo,Borschensky:2020hot,Dorsner:2018ynv,Mandal:2015lca}.

\begin{table}[t]
\renewcommand\arraystretch{1.2}
\centering
\begin{tabular}{|p{1cm}<{\centering} | p{2.cm}<{\centering} | p{2.5cm}<{\centering} | p{2.5cm}<{\centering} | p{2.5cm}<{\centering} | p{2.cm}<{\centering}|}
\hline
&   SM~\cite{MILC:2015uhg,Na:2015kha}                       & BaBar~\cite{BaBar:2012obs,BaBar:2013mob}                             & Belle$^\text{c}$~\cite{Belle:2019rba}                         & LHCb$^\text{a}$~\cite{LHCb:2023zxo}                          & HFLAV~\cite{HeavyFlavorAveragingGroupHFLAV:2024ctg}  \\
\hline
\multirow{2}*{$R(D)$}         & $0.298$                 & $0.440$                       & $0.307$                       & $0.441$                       & $0.342$                    \\
& $\pm 0.004 $          & $\pm 0.058 \pm 0.042$         & $\pm 0.037 \pm 0.016$         & $\pm 0.060 \pm 0.066$         & $\pm 0.026$                    \\
\hline
\multirow{2}*{$R(D^*)$}         & $ 0.254 $             & $0.332$                       & $0.283$                       & $0.281$                       & $0.287 $                     \\
& $\pm 0.005 $          & $\pm 0.024 \pm 0.018$         & $\pm 0.018 \pm 0.014$         & $\pm 0.018 \pm 0.024$         & $\pm 0.012$                     \\
\hline
\end{tabular}
\caption{The results of \( R(D^{(*)}) \) in SM, experimental collaborations and Heavy Flavor Averaging Group(HFLAV).}
\label{tab::Rds}
\end{table}

On the other hand, with the advancement of the Future Circular Collider (FCC-ee)~\cite{FCC:2018evy,FCC:2018byv,Blondel:2021ema,Agapov:2022bhm}, the Circular Electron Positron Collider (CEPC)~\cite{CEPCStudyGroup:2023quu,CEPCStudyGroup:2018ghi,CEPCStudyGroup:2018rmc}, and other proposed lepton colliders~\cite{CLICPhysicsWorkingGroup:2004qvu,CLICdp:2018cto,CLIC:2018fvx,LCCPhysicsWorkingGroup:2019fvj}, tests of the SM and searches for NP are expected to achieve unprecedented precision. 
One of the core objectives of these programs is to exploit the opportunities at the $Z$ pole by producing enormous numbers of $Z$ bosons, approximately $5\times10^{12}$ and $2.5\times10^{12}$ at the FCC-ee and the 30~MW CEPC, respectively~\cite{CEPCStudyGroup:2023quu,FCC:2018evy}. 
This allow determination of the $Z$ lineshape and the effective weak mixing angle $\sin^2\theta_W^{\rm eff}$ with a statistical uncertainty of a few per-mil or better at the FCC-ee~\cite{FCC:2018evy}. 
The process $e^+ e^- \to \mu^+ \mu^-$ at the $Z$ pole is a golden channel for a precise measurement of the forward--backward asymmetry $\mathcal{A}_{\text{FB}}^{ff}$, potentially reducing the uncertainty of $\sin^2\theta_W^{\rm eff}$ to about $6\times10^{-6}$~\cite{FCC:2018evy}. 
Meanwhile, the $\tau$ polarization asymmetry can similarly provide a comparably accurate for $\sin^2\theta_W^{\rm eff}$ measurement, with experimental precision better than $5\times10^{-6}$ expected at the FCC-ee~\cite{FCC:2018evy}. 
Such unprecedented precision allows us to probe possible deviations from the SM  at $Z$ factory, which may indicate the presence of NP.
Therefore, the production of $\mu$ and $\tau$-pair at the $Z$ factory plays a significant role in electroweak precision measurements.

Motivated by previous investigations, we calculate the effects of a scalar singlet LQ in $\tau$ and $\mu$-pair productions at the $Z$ factory.
The leading order (LO) processes are pure SM contribution, implying that the NP effects only appear at higher orders. 
Therefore, the NP effects are evaluated in those channels by calculating the next-to-leading order (NLO) corrections. 
This paper is organized as follows. 
In Section~\ref{sec::model}, we briefly introduce the Lagrangian of the scalar singlet LQ $S_1$. 
Focusing on the scenario addressing charged-current (CC) anomalies, a minimal framework is adopted, which only $\lambda^{1R}_{c\tau}$ and $\lambda^{1L}_{b\tau}$ are non-zero. 
The relevant experimental constraints that restrict the parameters of $S_1$ model are presented. 
In Section~\ref{sec::decay}, we explore the effects of $S_1$ in $Z \to\tau^+ \tau^-(\mu^+ \mu^-)$ decays. 
We first illustrate the dependence of the NP effects on the $S_1$ mass, followed by a parameter-space scan for the $\tau$-pair final state within the allowed region. 
Section~\ref{sec::sceI} presents the numerical results for $e^+e^-$ collision processes. 
The scalar $S_1$ exhibits the most significant effect near $\sqrt{s} = M_Z$ in $\tau$-pair production, with its dependence on the new couplings are identical to the case of $Z$ decay.
Further, we display the NP effects in a series of kinematic distributions. 
Finally, a brief summary is provided in Section~\ref{sec::sum}.

\section{Model}
\label{sec::model}

The Lagrangian of a scalar singlet LQ $S_1$ can be written as 
\begin{equation}\label{eq::lp} 
\begin{aligned}
\mathcal{L}_{\text{LQ}} &= |D_\mu S_1|^2 - M_{S_1}^2|S_1|^2 +\left[(\lambda^{1L})_{i\alpha}\bar{q}_i^c \epsilon l_\alpha + (\lambda^{1R})_{i\alpha}\bar{u}_i^c e_\alpha\right] S_1,
\end{aligned}
\end{equation}
where $\epsilon=i\sigma_2$, $D_\mu$ is the covariant derivative.
The $q_i$ and $l_\alpha$ fields stand for the SU(2)$_L$ doublets of left-handed quarks and leptons, 
whereas the $u_i$ and $e_\alpha$ stand for the SU(2)$_L$ singlets of up-type quarks and charged leptons.
The couplings $\lambda^{1R(L)}$ can be parametered as 
\begin{equation}
\begin{aligned}
\lambda^{1R}=
\begin{pmatrix}
\lambda^{1R}_{ue} & \lambda^{1R}_{u\mu} & \lambda^{1R}_{u\tau} \\
\lambda^{1R}_{ce} & \lambda^{1R}_{c\mu} & \lambda^{1R}_{c\tau} \\ 
\lambda^{1R}_{te} & \lambda^{1R}_{t\mu} & \lambda^{1R}_{t\tau} \\
\end{pmatrix},\qquad
\lambda^{1L}=
\begin{pmatrix}
\lambda^{1L}_{de} & \lambda^{1L}_{d\mu} & \lambda^{1L}_{d\tau} \\
\lambda^{1L}_{se} & \lambda^{1L}_{s\mu} & \lambda^{1L}_{s\tau} \\ 
\lambda^{1L}_{be} & \lambda^{1L}_{b\mu} & \lambda^{1L}_{b\tau} \\
\end{pmatrix}.
\end{aligned}
\end{equation}

\begin{figure}[t]
\centering
\includegraphics[scale=0.6]{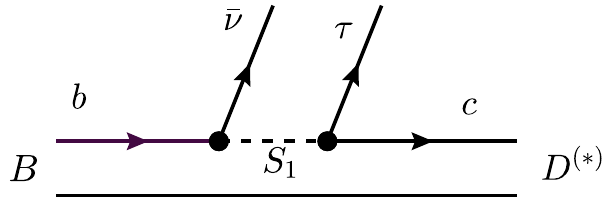}
\caption{The contribution of single scalar LQ $S_1$ to semi-leptonic B-meson decay is shown.}
\label{fig::LQ}
\end{figure}

In this article, we focus on the scenario where $S_1$ addresses the CC anomalies, i.e. the deviations in the \( b \to c \tau \nu \) transitions, as shown in Fig.~\ref{fig::LQ}. 
Two non-zero couplings ($\lambda^{1R}_{c\tau}$ and $\lambda^{1L}_{b\tau}$) with a TeV-scale LQ are sufficient to explain the CC anomalies, and even simultaneously accommodate the neutral-current anomalies. 
The collider searches at the LHC set lower bounds on LQ masses, approaching the $\approx$ 1 TeV~\cite{Marzocca:2018wcf,Saad:2020ihm}. 
In this work, we work within a minimal framework in which only $\lambda^{1R}_{c\tau}$ and $\lambda^{1L}_{b\tau}$ are non-zero, and present results with $M_{S_1}=$ 1 TeV and 2 TeV.  
In Fig.~\ref{fig::scandecay}, the $2\sigma$ limits from the $R(D^{(*)})$ measurements are shown as blue lines in the ($\lambda^{1R}_{c\tau}$, $\lambda^{1L}_{b\tau}$) plane. 
The constraint from the measurement of 
$\mathrm{Br}(B_c^+ \to \tau^+ \nu)$~\cite{Akeroyd:2017mhr,Gherardi:2020qhc} is shown as the purple line in FIG.~\ref{fig::scandecay}.
Meanwhile, the introduction of the left-handed coupling $\lambda^{1L}_{b\tau}$ induces the loop corrections to the $W$-$\tau$-$\nu_\tau$ vertex.
In this work, following Ref.~\cite{Gherardi:2020qhc}, the ratio $|g_{\tau}/g_\mu|$ is used to test the flavour universality of the $W$ coupling as
\begin{equation}
\begin{aligned}
%|g_{\tau}/g_\mu|^2 &= \frac{R_{\Gamma_{\tau e}}}{R_{\Gamma_{\mu e}}} = 1-\frac{1}{1000}\frac{(\lambda_{b\tau}^{1L})^2}{m_{1}^2}(1.61 + 0.67 \log m_1^2),\\ 
|g_{\tau}/g_\mu|^2 &= \frac{R_{\Gamma_{\tau e}}}{R_{\Gamma_{\mu e}}} = 1-\frac{1}{1000}\frac{(\lambda_{b\tau}^{1L})^2}{(M_{S_1}/\text{TeV})^2}\left[1.61 + 0.67 \log (M_{S_1}/\text{TeV})^2\right].\\ 
\end{aligned}
\end{equation}
%where $m_1=\frac{M_{S_1}}{\rm{TeV}}$ denotes the $S_1$ mass in units of TeV.
This ratio can be extracted from various decay channels including the purely leptonic 
$\tau$ and $\mu$ decays, the leptonic and semi-leptonic $\pi/K$ decays~\cite{HeavyFlavorAveragingGroupHFLAV:2024ctg}, and the direct leptonic decays of the $W$ boson~\cite{ParticleDataGroup:2026aaa}. 
Following Ref.~\cite{Pich:2025goe}, we adopt the average result of $|g_{\tau}/g_\mu|=1.0006 \pm 0.0013$ obtained by combining the above measurements.
The collider constraints are analyzed using the {\tt HighPT} package~\cite{Allwicher:2022mcg,Allwicher:2022gkm}. 
The upper and lower panels of FIG.~\ref{fig::scandecay} correspond to $M_{S_1}=1$~TeV and $2$~TeV, respectively. 
In following, two benchmark points (BPs) listed in Table.~\ref{tab::BPs1} are chosen to perform our numerical analysis.

\begin{table}[t]
\renewcommand\arraystretch{1.5}
\centering
\begin{tabular}{| p{1cm}<{\centering} |p{2cm}<{\centering} |p{2cm}<{\centering} |p{2cm}<{\centering} |}
\hline
& $M_{S_1}$(GeV)       & $\lambda_{b\tau}^{1L}$     & $\lambda_{c\tau}^{1R}$    \\
\hline
BP0     &   1000                &  1.4                      &   -0.1                    \\
\hline
BP1     &   2000                &  2.3                      &   -0.4                    \\
\hline
\end{tabular}
\caption{
The BPs chosen to perform the numerical analysis.
The BP0 and BP1 correspond to the black points in upper and lower panels of FIG.~\ref{fig::scandecay}, respectively.
}
\label{tab::BPs1}
\end{table}

To quantify the NP effects originating from $S_1$ at the $Z$ factory, following Ref.~\cite{He:2024bwh}, we define the parameter $\delta$ as
\begin{equation}\label{eq::delta}
\begin{aligned}
\delta = \frac{\sigma^{\text{NLO}}_{S_1} - \sigma^{\text{NLO}}_{\text{SM}}}{\sigma^{\text{LO}}} = \frac{\Delta \sigma^{\text{NLO}}_{S_1}}{\sigma^{\text{LO}}},
\end{aligned}
\end{equation}
where the $\sigma^{\text{NLO}}_{S_1(\text{SM})}$ are the cross sections in LQ and SM models at NLO, respectively.
For the case of $Z$ decay, the cross section $\sigma$ will be modified to the decay width $\Gamma$.
Accordingly, the $\Delta \sigma^{\text{NLO}}_{S_1}(\Delta \Gamma^{\text{NLO}}_{S_1})$ denote the pure $S_1$ contributions at NLO.
Since the precision at next-generation machines is expected to be extremely good, the value of $\delta$ at the level of 0.001(0.1\%) or better to be reached in our study.

The calculations of NLO cross section or decay width adhere to the framework of {\tt SloopS}~\cite{Boudjema:2005hb,Baro:2007em,Baro:2008bg,Boudjema:2009pw,Baro:2009na,Boudjema:2011ig,Boudjema:2014gza,Belanger:2016tqb,Belanger:2017rgu,Banerjee:2019luv,He:2024bwh},
in which the NLO model is built by {\tt Lanhep}~\cite{Semenov:2014rea,Semenov:2010qt,Semenov:2008jy,Semenov:2002jw,Semenov:1998eb,Semenov:1996es},
the generation and integration of squared amplitude are performed by {\tt FeynArts} \cite{Kublbeck:1990xc}, {\tt  FormCalc} \cite{Mertig:1990an} and {\tt Looptools} \cite{Hahn:1998yk} ({\tt FFL}).

\section{The $\tau$ and $\mu$-pair productions in $Z$ decay}
\label{sec::decay}

\begin{figure}[ht]
\centering
\includegraphics[scale = 1]{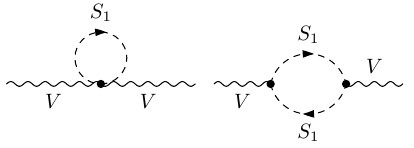}
\includegraphics[scale = 1]{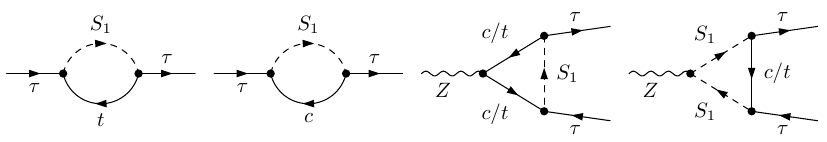}
\caption{
The contributions of $S_1$ to V-bosons self-energy functions, $\tau$ lepton self-energy function and $Z\tau \tau$ vertex, where $V=Z$ and $\gamma$.
}
\label{fig::zdecaynlo}
\end{figure}

At LO, the $Z \to \tata$ decay is a pure SM process. 
Consequently, the NP effects from $S_1$ are indirect and appear only in higher order corrections. 
In other words, the NLO EW corrections are essential for probing the $S_1$ effects at the $Z$ factory.
In the CC scenario, the $S_1$ could modify the self-energy functions of the vector bosons, and further contributes to the renormalization constants of the $Z$ boson wave function and the electric charge, as shown in the upper panel of FIG.~\ref{fig::zdecaynlo}.
However, the introduction of $\lambda^{1R}_{c\tau}$ and $\lambda^{1L}_{b\tau}$ couplings induces the self-energy corrections of $\tau$ lepton and the $Z\tau \tau$ vertex corrections for $\tau$-pair production, as depicted in the lower panel of FIG.~\ref{fig::zdecaynlo}. 
The latter effects would be sensitive to the top quark mass by introducing the internal top quark in triangle loop.

\begin{figure}[t]
\centering
\includegraphics[width = 0.49\textwidth, height = 5.8cm]{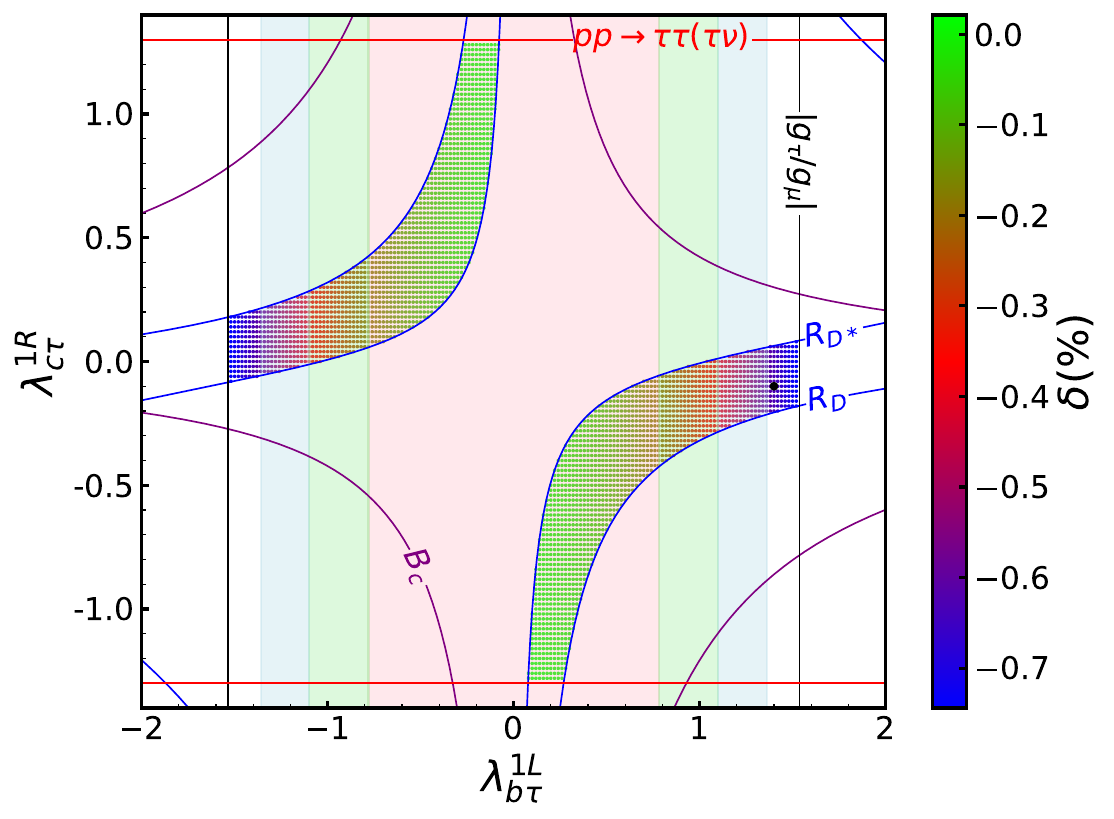}
\includegraphics[width = 0.49\textwidth, height = 5.8cm]{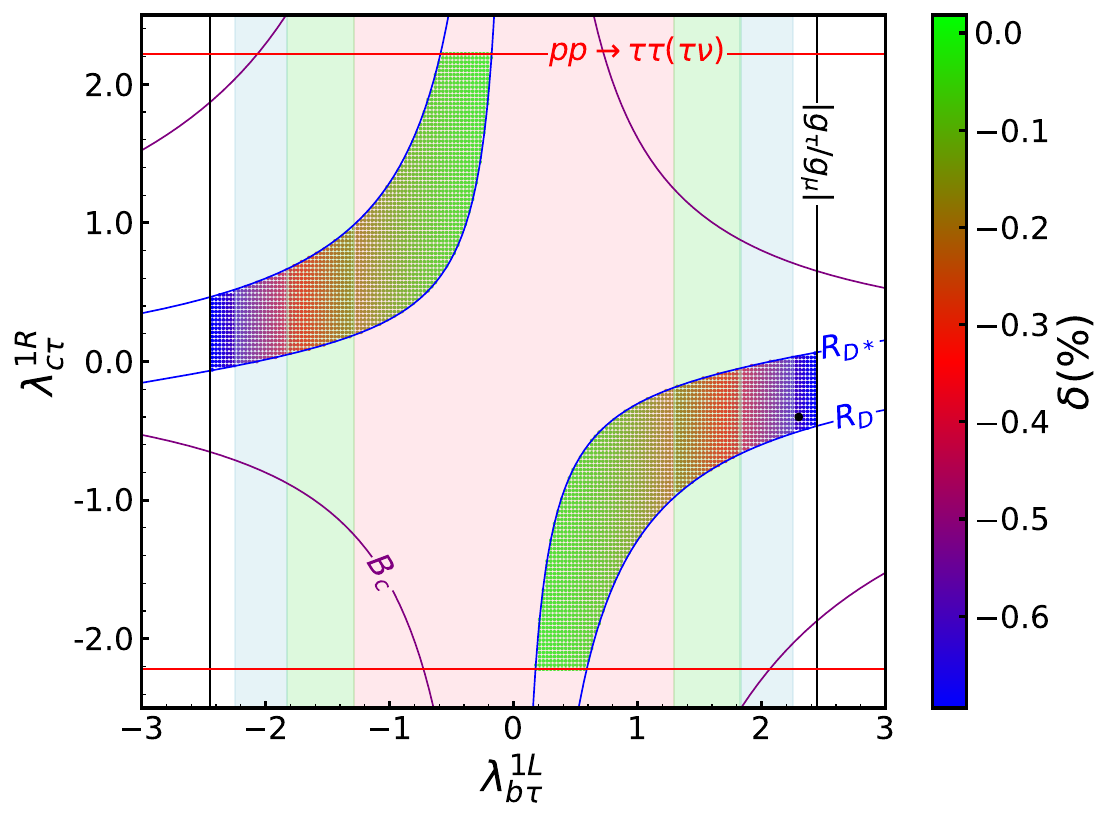}
\caption{The contributions of $S_1$ to $Z \to \tau^+ \tau^-$ decay.
The left and right panels are the results of $M_{S_1}=1,2$ TeV, respectively.
}
\label{fig::scandecay}
\end{figure}

For the $Z \to \tau^+ \tau^-$ decay, we scan the NP effects in the ($\lambda_{b\tau}^{1L}, \lambda_{c\tau}^{1R}$) parameter space, as shown in FIG.~\ref{fig::scandecay}. 
The left and right panels correspond to the results of $M_{S_1}=1,2$ TeV in allowed parameters space, respectively.
Moreover, these results mainly focus on the regions of $(\lambda_{b\tau}^{1L}>0, \lambda_{c\tau}^{1R}<0)$ and $(\lambda_{b\tau}^{1L}<0, \lambda_{c\tau}^{1R}>0)$, which are approximately symmetric with $\lambda_{b\tau}^{1L} = \lambda_{c\tau}^{1R} = 0$. 
The NP effects are almost independent on the right-handed coupling $\lambda_{c\tau}^{1R}$, but become more pronounced as the absolute value of $\lambda_{b\tau}^{1L}$ increases. 
The $S_1$ effects are mainly constrained by the $|g_\tau/g_\mu|$ measurements ($\lambda_{b\tau}^{1L} \lesssim 1.54$ for $M_{S_1}=1$~TeV), with a maximum deviation of about $-0.7\%$. 
The results for $M_{S_1}=2$~TeV are similar, but the allowed parameter space is significantly broader. 
The upper bound on $\lambda_{b\tau}^{1L}$ from the $|g_\tau/g_\mu|$ measurement extends to approximately $2.45$ for $M_{S_1}=2$~TeV, while the most pronounced NP effect remains around $-0.7\%$. 
Comparing the results of $M_{S_1}=1$ and $2$~TeV, an important point is that the suppression of NP effects due to the heavy $S_1$ mass can be compensated by the expanded allowed parameter space.

The non-SM contributions in $Z \to \tau^+ \tau^-$ decay from current $Z$-resonance experiments~\cite{ALEPH:2005ab} are more relaxed than the current $|g_\tau/g_\mu|$ constraints at 95\% CL. 
Furthermore, the anticipated measurement precision at the future $Z$ factory is expected to impose stricter constraints on the $S_1$ model, particularly on the left-handed coupling $\lambda_{b\tau}^{1L}$. 
In FIG.~\ref{fig::scandecay}, the anticipated precisions of the future $Z$ factory are mapped to the ($\lambda_{b\tau}^{1L}, \lambda_{c\tau}^{1R}$) plane. 
The pink, green, and blue regions correspond to measurement precisions of 0.1\%, 0.2\% and 0.3\% at 95\% CL, respectively. 
Those conclusions also apply to the $e^+ e^- \to \tau^+ \tau^-$ process, which will be discussed later.

\begin{figure}[t]
\centering
\includegraphics[width = 0.48\textwidth, height = 5.54cm]{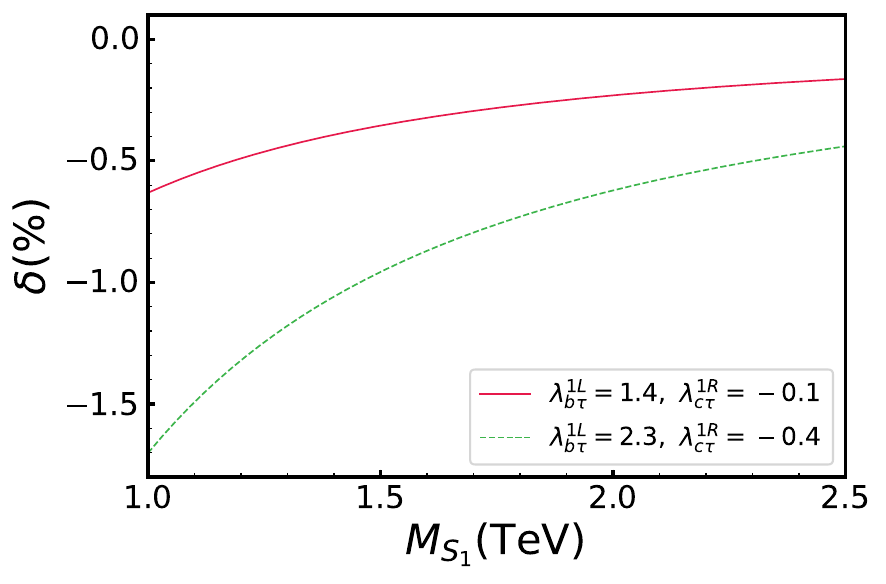}
\includegraphics[width = 0.48\textwidth, height = 5.8cm]{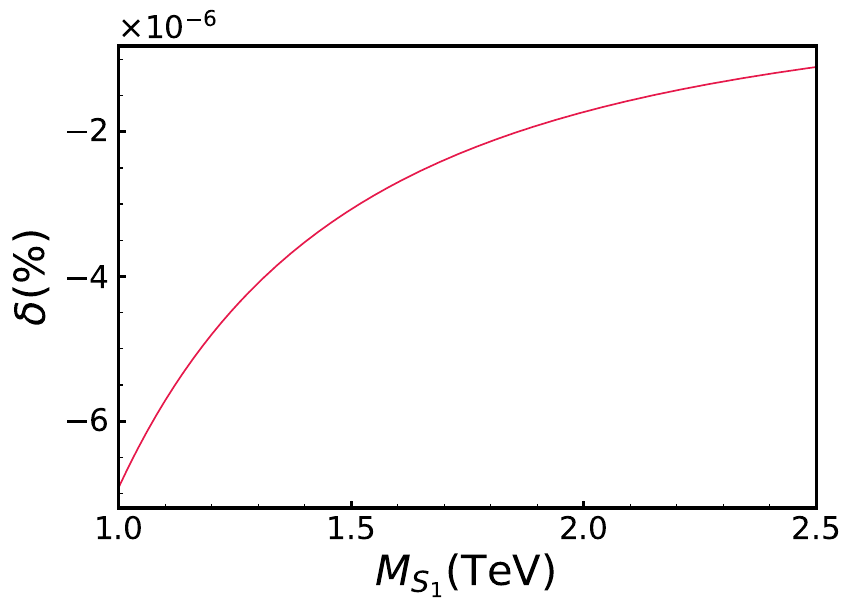}
\caption{
The dependence of NP effects on $M_{S_1}$ for $Z \to \tau^+ \tau^-$ and $Z \to \mu^+ \mu^-$ decays are shown in left and right panels. 
The couplings in the left panel fixed to the values of BP0 and BP1.
}
\label{fig::decay_ms_llcc}
\end{figure}

\begin{figure}[t]
\centering
\includegraphics[width = 0.49\textwidth, height = 5.8cm]{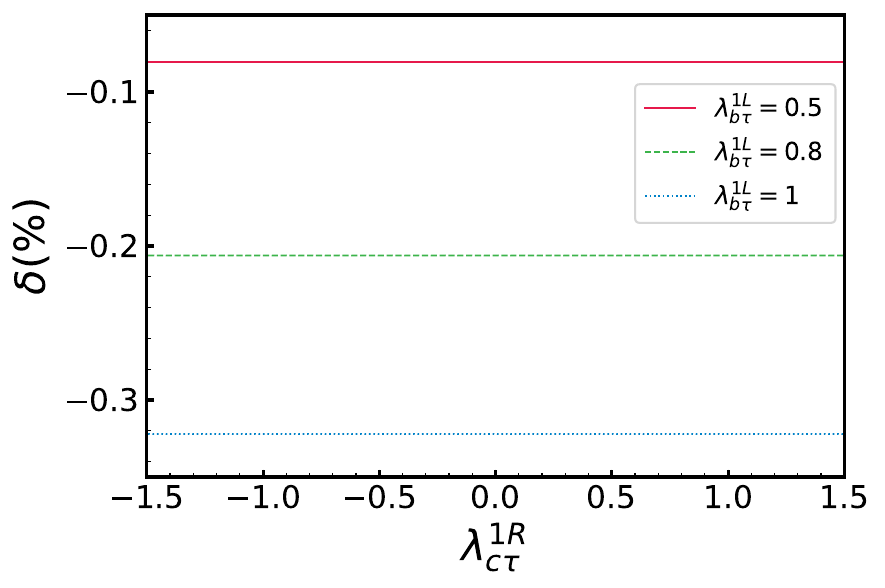}
\includegraphics[width = 0.49\textwidth, height = 5.8cm]{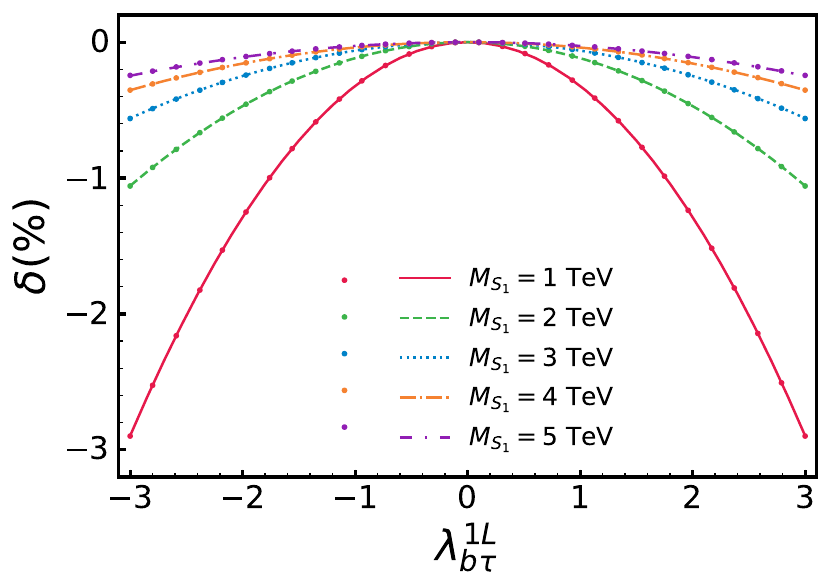}
\caption{
The left panel shows the dependence of NP effects on the right-hand coupling by fixing the $\lambda_{b\tau}^{1L}=$ 0.5, 0.8, 1 and $M_{S_1}=$ 1 TeV. 
The right panel shows the dependence of the NP effects on the left-handed coupling $\lambda_{b\tau}^{1L}$, with $\lambda_{c\tau}^{1R}$ fixed to zero. 
The numerical results we calculated are indicated by the dots, while the curves correspond to the results obtained from function Eq.~\ref{eq:fitted}.
}
\label{fig::fitted}
\end{figure}

To further illustrate our results, two BPs based on the results of FIG.~\ref{fig::scandecay} are chosen, which listed in Table.~\ref{tab::BPs1}.
In the left panel of FIG.~\ref{fig::decay_ms_llcc}, we show the result of $S_1$ contributions as a function of $M_{S_1}$ for the $Z \to \tau^+ \tau^-$ decay, with the couplings fixed at the values in Table~\ref{tab::BPs1}. 
It can be seen that the contributions from the virtual $S_1$ in the loop reduce the SM decay rate. 
With the increment of the $S_1$ mass, the reduction effects decrease.
Based on the results of FIG.~\ref{fig::scandecay}, we further exhibit the independence of NP effects on the right-hand coupling by fixing the left-hand coupling $\lambda_{b\tau}^{1L}$ in the left panel of FIG.~\ref{fig::fitted}.  
The $S_1$ contributions are almost entirely independent of right-hand coupling.
Meanwhile, the dependence of the NP effects on the left-handed coupling can be more clearly illustrated by fixing the right-handed coupling.  
As shown in the right panel of FIG.~\ref{fig::fitted}, the differently colored data points indicate that the NP effect exhibits a clear quadratic dependence on the left-handed coupling ($\lambda_{b\tau}^{1L}$) with $\lambda_{c\tau}^{1R}=0$ fixed, and this behavior holds for different $S_1$ masses. 
Combining the results in FIG.~\ref{fig::decay_ms_llcc}, we fit a analytic function $\delta_{\text{fitted}} $ to describe the effects of $S_1$ in $Z \to \tau^+ \tau^-$ decay, which is given in Eq.~\ref{eq:fitted}.
\begin{equation}\label{eq:fitted}
\begin{aligned}
\delta_{\text{fitted}} (\lambda_{b \tau}^{1L}, M_{S_1}) =  (\lambda_{b \tau}^{1L})^2 \times \left[ \frac{K_2}{(M_{S_1} + K_d)^2} + \frac{K_1}{M_{S_1} + K_d} \right], 
\end{aligned}
\end{equation}
where $K_2 = -0.5919,\ K_1 = -0.03947$ and $K_d = 0.4188$.
In FIG.~\ref{fig::fitted}, the predictions of the fitted function $\delta_{\text{fitted}} $ for different $S_1$ masses are shown as distinctive curves.

In the scenario addressing the CC anomalies, there is no coupling between 2th-generation leptons and quarks.
Therefore, for $Z \to \mu^+ \mu^-$ decay, the $S_1$ only alters the self-energy functions of the vector bosons, as shown in the upper panel of FIG.~\ref{fig::zdecaynlo}, and further contributes to the corresponding counter terms. 
Obviously, the NP effects from $S_1$ in $Z \to \mu^+ \mu^-$ are independent on the couplings $\lambda_{b\tau}^{1L}$ and $\lambda_{c\tau}^{1R}$.
The behavior of the $S_1$ contribution as a function of $M_{S_1}$ is similar to that in the $Z \to \tau^+ \tau^-$ channel, as shown in the right panel of FIG.~\ref{fig::decay_ms_llcc}, but its magnitude is approximately $\mathcal{O}(10^{-6})\%$.
Such NP effects in the $Z \to \mu^+ \mu^-$ channel are essentially negligible, also implying that the current measurements do not provide further constrain on the mass of the scalar singlet LQ.
We have also evaluated the corresponding corrections to $Z\to b\bar b$, $Z\to c\bar c$ and $Z\to\nu_\tau\bar\nu_\tau$ channels.
The detailed results are collected in Appendix~\ref{appA}.
We find that these additional $Z$-pole channels do not provide stronger sensitivities than the $\tau^+\tau^-$ channel in the present minimal scenario.

\section{The $\tau$ and $\mu$-pair productions at $e^+ e^-$ collider}
\label{sec::sceI}

\begin{figure}[H]
\centering
\includegraphics[scale = 1]{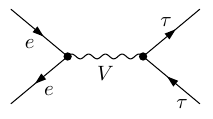}
\includegraphics[scale = 1]{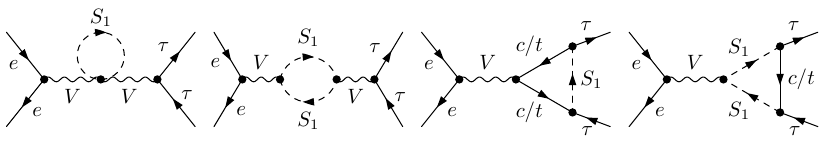}
\caption{The Feynman diagrams at LO and the loop corrections induced by $S_1$ for the $e^+ e^- \to \tau^+ \tau^-$ process are shown in the upper and lower panels, respectively.}
\label{fig::eefflo}
\end{figure}

\begin{figure}[t]
\centering
\includegraphics[scale = 0.49]{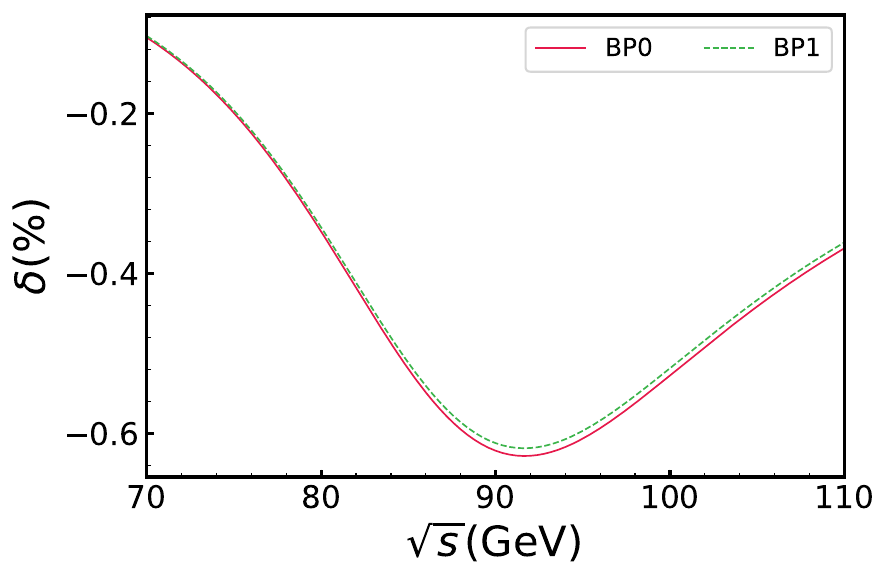}
\includegraphics[scale = 0.49]{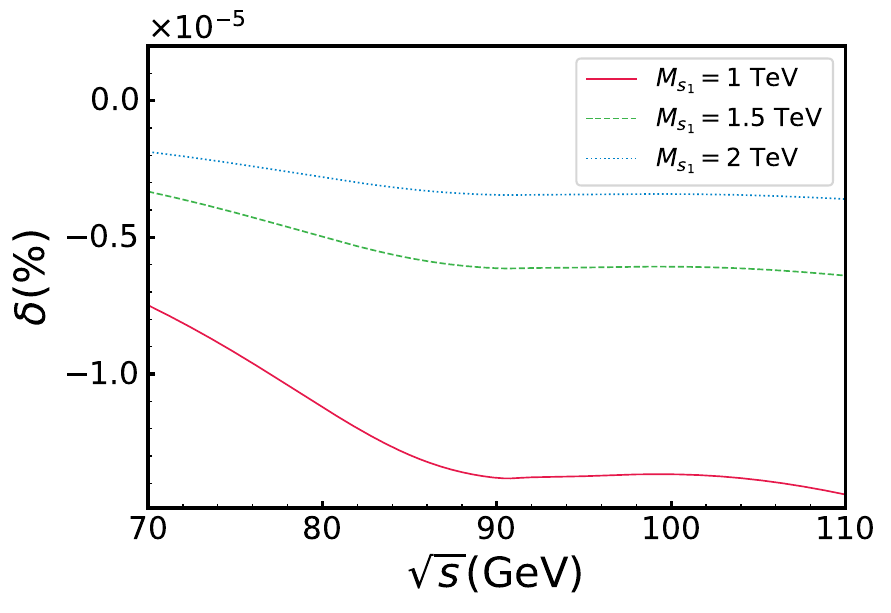}
\caption{
The dependence of NP effects($\delta$) on the $M_{S_1}$ and collision energy $\sqrt{s}$.
The left and right panels are the results of $\tata$ and $\mumu$ productions, respectively. 
}
\label{fig::ms-sqrts_mmcc}
\end{figure}

\begin{table}[t]
\renewcommand\arraystretch{1.3}
\centering
\begin{tabular}{|p{1.7cm}<{\centering} |p{2.cm}<{\centering} |p{2.5cm}<{\centering} |p{1.8cm}<{\centering} |p{2.5cm}<{\centering} |p{1.8cm}<{\centering} |p{2cm}<{\centering} |p{2cm}<{\centering} |p{2cm}<{\centering} |}
\hline
$\sqrt{s}$ (GeV)  & $\sigma^{\text{LO}}$ (pb)& $\Delta \sigma^{\text{NLO}}_{S_1\ \text{BP0}}$ (pb)  & $\delta_{\text{BP0}}$  & $\Delta \sigma^{\text{NLO}}_{S_1\ \text{BP1}}$ (pb)  & $\delta_{\text{BP1}}$\\
\hline
88.231 & 281.91(9)   &   -1.6889(5)    &  -0.60\%   &  -1.6634(5)     &  -0.59\%   \\
\hline
89.236 & 540.2(1)   & -3.3161(9)    &  -0.61\%   &  -3.2657(9)     &  -0.60\%   \\
\hline
90.238 & 1199.5(2)  & -7.477(2)     &  -0.62\%   &  -7.363(2)      &  -0.61\%   \\
\hline
91.230 & 1956.8(3)  & -12.283(2)    &  -0.63\%   &  -12.095(2)     &  -0.62\%   \\
\hline
92.226 & 1159.7(1)  & -7.2769(7)    &  -0.62\%   &  -7.1639(7)     &  -0.62\%   \\
\hline
93.228 & 537.5(1)  & -3.3483(1)    &  -0.61\%   &  -3.2958(1)     &  -0.61\%   \\
\hline
94.223 & 291.19(5) & -1.79033(3)   &  -0.60\%   &  -1.76199(3)    &  -0.61\%   \\
\hline
\end{tabular}
\caption{
The $S_1$ contributions for the center of mass energies measured by LEP near the $Z$ pole are presented.
$\Delta \sigma^{\text{NLO}}_{S_1}$ denotes the contribution to the cross section from $S_1$, coming from Eq.~\ref{eq::delta}.
}
\label{tab::reslep}
\end{table}

The LO Feynman diagram for the $e^+ e^- \to \tau^+ \tau^- $ is shown in the upper of FIG.~\ref{fig::eefflo}. 
As well, the LO process is governed by SM contributions, implying that the NP effects are indirect and detectable only at higher order corrections. 
The interactions associating with 1-th generation fermions are not considered, which prevents $S_1$ from contributing to the $Vee$ vertex by loop diagrams.
Therefore, for $\sqrt{s}=M_Z$, we have
\begin{equation}\label{eq:relation}
\begin{aligned}
& \sigma_{\text{peak}}^{\text{LO}}=12 \pi \frac{\Gamma(Z \to ee) \Gamma(Z \to f \bar f)}{M_Z^2 \Gamma_Z^2}, \qquad f=\mu, \tau, \\
& \frac{\Delta \sigma_{S_1}^{\text{NLO}}( e^+ e^- \to f \bar f)}{\sigma^{\text{LO}}( e^+ e^- \to f \bar f)}\bigg|_{\sqrt{s}=M_Z}\approx \frac{\Delta \Gamma_{S_1}^{\text{NLO}}(Z \to f \bar f)}{\Gamma^{\text{LO}}(Z \to f \bar f)} 
\end{aligned}
\end{equation}
The NP effects at $e^+ e^-$ collider at $\sqrt{s}=M_Z$ should be consistent with the case of $Z$ decay.
The $S_1$ contributions to $\tau$-pair productions are shown in the lower panel of FIG.~\ref{fig::eefflo}, respectively. 
The dependence of NP effects on the $\sqrt{s}$ for $\tau$-pair production is shown in the left panel of  FIG.~\ref{fig::ms-sqrts_mmcc}, where the two BPs exhibit similar behavior. 
For $\sqrt{s} < M_Z$, the $S_1$ contributions increase with increasing $\sqrt{s}$, reaching a maximum of about $-0.65\%$ at $\sqrt{s} = M_Z$. 
For $\sqrt{s} > M_Z$, the value of $\delta$ decreases as $\sqrt{s}$ increases. 
Meanwhile, the results for the center of mass energies measured by LEP near the $Z$ pole are presented in Table~\ref{tab::reslep}.  
The contribution from $S_1$ will be reduced to below $0.1\%$, making it unobservable, for the further extended range of $\sqrt{s} \sim 130$ – $207~\mathrm{GeV}$ at the LEP.
As we mentioned, the scan of NP effects in the $(\lambda_{b\tau}^{1L},\lambda_{c\tau}^{1R})$ plane is identical with the decay  (FIG.~\ref{fig::scandecay} in Sec.~\ref{sec::decay}), we will not elaborate further.

\begin{figure}[t]
\centering
\includegraphics[width = 0.49\textwidth, height = 6.5cm]{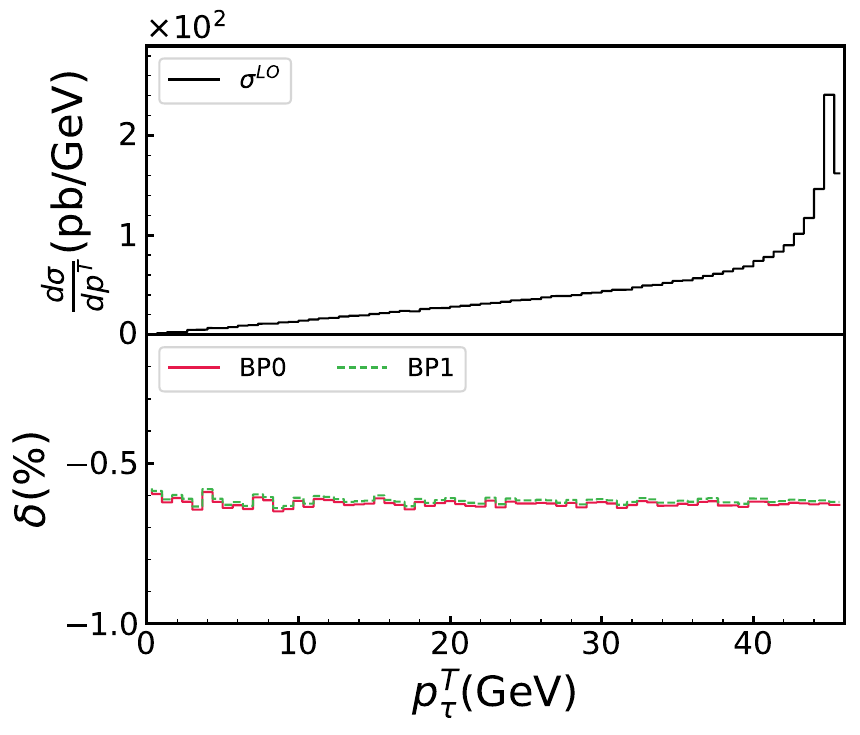}
\includegraphics[width = 0.49\textwidth, height = 6.5cm]{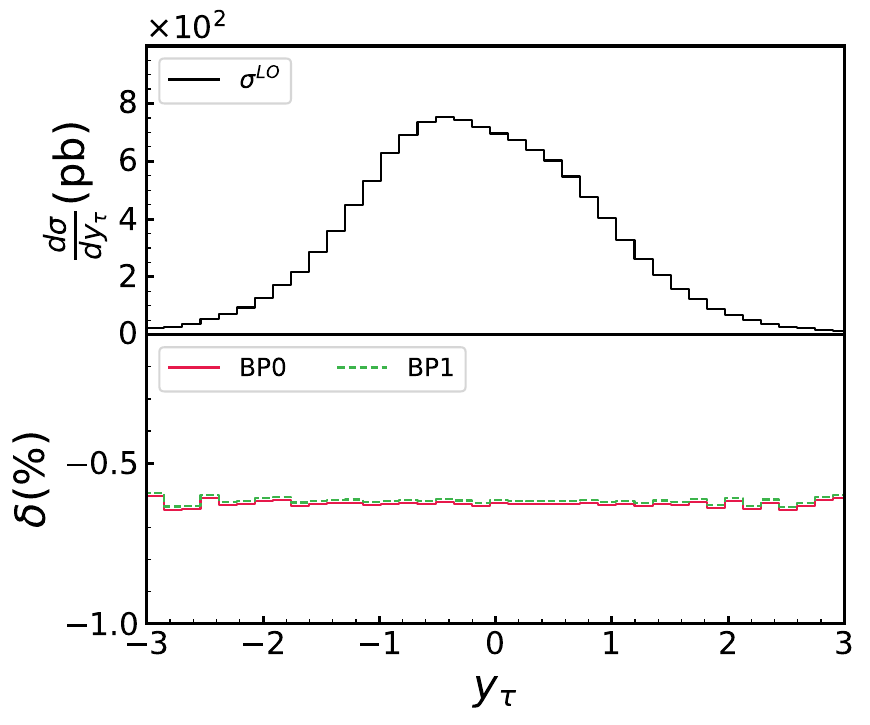}
\includegraphics[width = 0.49\textwidth, height = 6.5cm]{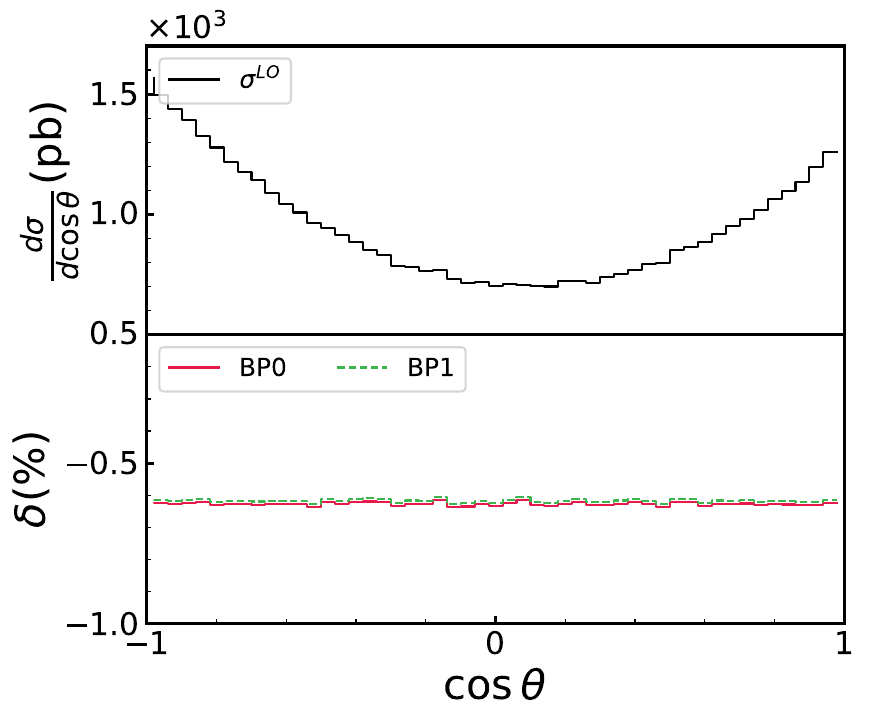}
\caption{
The various kinematic distributions in $e^+ e^- \to \tata$ process are presented.
The results from top to bottom are the distributions of transverse momentum, rapidity, angular of final $\tau^+$, respectively.
}
\label{fig::dis1}
\end{figure}

The $\tau$-pair production channel is a promising candidate for exploring $S_1$ effects near the $Z$ pole. 
Various kinematic distributions in $\tau$-pair production at the $Z$ pole are presented in FIG.~\ref{fig::dis1}. 
The results from top to bottom in FIG.~\ref{fig::dis1} are the distributions of transverse momentum, rapidity and angular of final state $\tau^-$.
In the upper panels of all subfigures, we present the distributions of kinematic variables at LO. 
Since the inclusion of NP effects does not alter their behavior compared to the LO results, they are not shown here.
The results indicate that the final states are predominantly concentrate on specific kinematic regions (e.g., $p_\tau^T > 30$~GeV, $|y_\tau| < 2$, and $|\cos\theta| > 0.5$).
The lower panels show that the NP effects are independent of the kinematic variables, and remain stable across each bins.
Moreover, the deviations observed in the differential distributions are consistent with the corresponding corrections to the total cross section.

Analogous to the case of $Z \to \mumu$, the $S_1$ only contributes to the self energy functions of  internal $\gamma/Z$ boson for $e^+ e^- \to \mumu$, which suppressed by the large $S_1$ mass and independent on the new couplings. 
In the right panel of FIG.~\ref{fig::ms-sqrts_mmcc}, we show the dependence of NP effects on the $S_1$ mass and collision energy. 
To better illustrate the dependence on $M_{S_1}$, additional results for $M_{S_1}=1.5$~TeV are included. 
The results are suppressed by the $S_1$ mass. 
And the behavior of $\delta$ as a function of $\sqrt{s}$ shows that it increases significantly with increasing $\sqrt{s}$ and stabilizes for $\sqrt{s} > M_Z$. 
Consequently, these NP effects are practically indistinguishable from the SM in $\mu$-pair production.

\section{Summary}
\label{sec::sum}
In this work, we investigate the NP effects of a scalar singlet LQ $S_1$ at the $Z$ factory by calculating the NLO corrections. 
In the CC scenario, the contributions of $S_1$ to vector-boson self-energy corrections are negligible, and the deviations from SM in the $\mu$-pair production channels are undetectable.  
For the $\tau$-pair production, the scans of $(\lambda_{b\tau}^{1L}, \lambda_{c\tau}^{1R})$ parameter space indicate that the $S_1$ contributions are primarily sensitive to the left-handed coupling $\lambda_{b\tau}^{1L}$. 
The maximal NP effect reaches approximately $-0.7\%$ for both $M_{S_1} = 1$ and $2$~TeV. 
An interesting point is that the suppression of NP effects due to the heavy $S_1$ mass can be compensated by the enlarged allowed parameter space. 
An analytical expression is given to describe the $S_1$ effects in the $\tau$-pair production.
We also provide the coupling constraints corresponding to the current and expected measurement precisions at the $Z$ factory.
In the collision processes, the $S_1$ contributions are most pronounced at the $Z$ pole. 
The kinematic distributions demonstrate that the NP contributions remain stable across bins of dynamical variables.
Moreover, precision observables at the future $Z$ factory, such as the forward-backward asymmetry and the effective weak mixing angle, are expected to receive measurable contributions from $S_1$, and show deviations from SM expectations.

\acknowledgments
D.Z.H. is supported by the Heze University doctoral fund project No.~010008002039037.
Y.Z. is supported by the National Natural Science Foundation of China under Grant No.~12475106 
     and the Fundamental Research Funds for Central Universities under Grant No.~JZ2023HGTB0222. 
%X.L. is supported by the National Natural Science Foundation of China under Grant No.~12205002.
H.S. is supported by the National Natural Science Foundation of China under Grant No.~12075043. 
%(rong lin)
%This work is supported by the Shandong Province Higher Education Institutions Youth Innovation Team Program (2023KJ279) and Heze University doctoral fund project (XY22BS30 and No.~010008002039037)。

\appendix
\section{Results for $Z \to f \bar f (f=c,b$ and $\nu_\tau$) channels}
\label{appA}

\begin{figure}[H]
\centering
\includegraphics[scale=0.7]{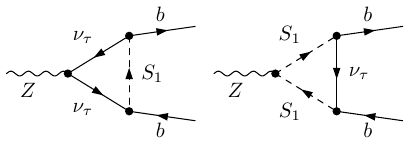}
\includegraphics[scale=0.7]{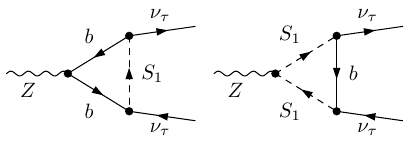}
\includegraphics[scale=0.7]{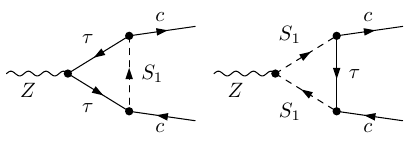}
\caption{
The Feynman diagrams of $S_1$ contributions to $Z \to f \bar f (f=c,b$ and $\nu_\tau$) channels are shown.
}
\label{fig::decayll}
\end{figure}

\begin{figure}[H]
\centering
\includegraphics[width = 0.49\textwidth, height = 5.8cm]{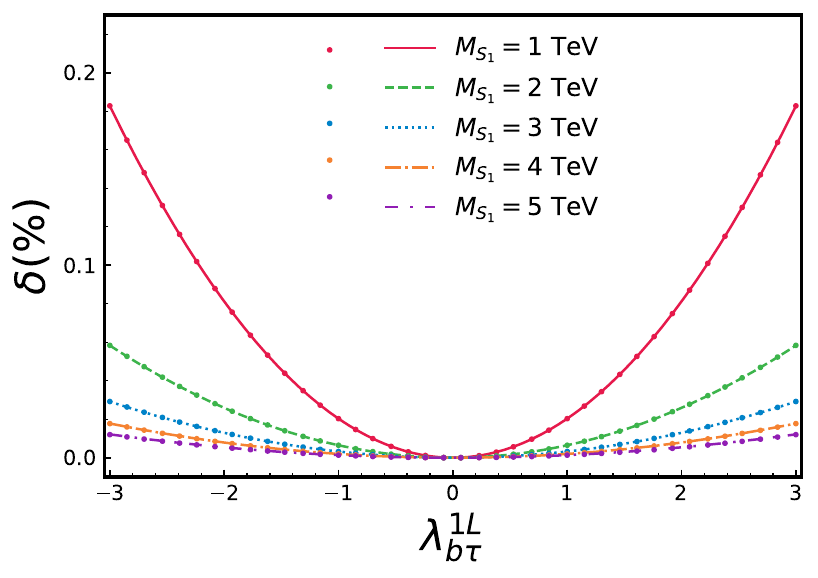}
\includegraphics[width = 0.49\textwidth, height = 5.8cm]{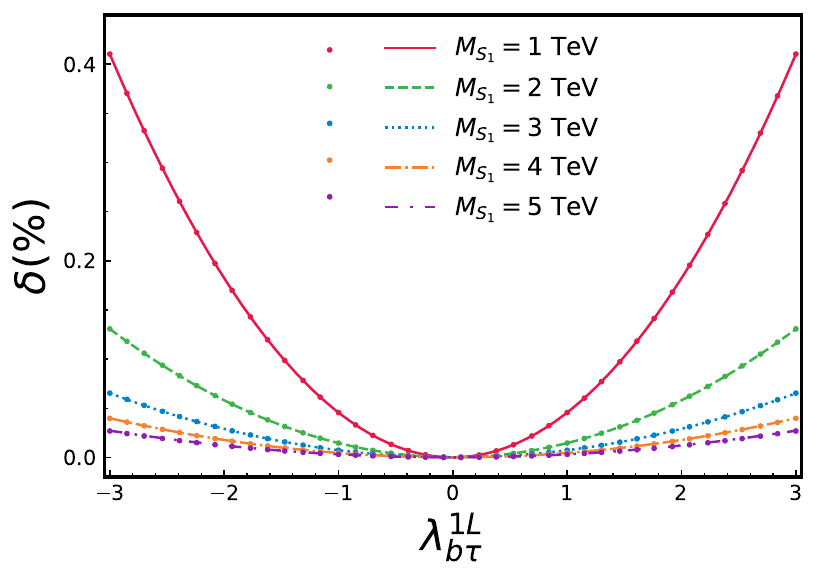}
\caption{
The results of $Z \to b \bar b$ and $Z \to \nu_\tau \bar{\nu}_\tau$ decays are shown in left and right panels, respectively. 
The results of lines are calculated from the functions of Eq.~\ref{eq:fitted}.
}
\label{fig::fittedz}
\end{figure}
\begin{table}[h]
\renewcommand\arraystretch{1.3}
\centering
\begin{tabular}{| p{2cm}<{\centering} |p{2.3cm}<{\centering} |p{2.3cm}<{\centering} |p{2.3cm}<{\centering} |}
\hline
$Z \to f \bar f$ & $K_2$       & $K_1$     & $K_d$    \\
\hline
$Z \to \tau^+ \tau^-$ & -0.5919 &       -0.03947 &     0.4188    \\
\hline
$Z \to b \bar b$ &     0.02857 &    0.001587 &     0.2260    \\
\hline
$Z \to \nu_\tau\bar\nu_\tau$ &      0.06379 &       0.003537 &      0.2226    \\
\hline
\end{tabular}
\caption{
The coefficients of fitted function $\delta_{\text{fitted}}$ Eq.~\ref{eq:fitted} for different decay channels. 
}
\label{tab::coefs}
\end{table}

In this appendix, we present the $S_1$ induced contributions to $Z \to f \bar f (f=c,b$ and $\nu_\tau$) in the present minimal $S_1$ model. 
The Feynman diagrams are shown in FIG.~\ref{fig::decayll}.
The corrections from $S_1$ to $Z\to b\bar b$ and $Z\to\nu_\tau\bar\nu_\tau$ are only dependent on the left-handed coupling $\lambda^{1L}_{b\tau}$, whereas the correction to $Z\to c\bar c$ is controlled by the right-handed coupling $\lambda^{1R}_{c\tau}$.
We also provide the fitted function $\delta_{\text{fitted}}$ Eq.~\ref{eq:fitted} to describe the NP effects for the $b\bar b$ and $\nu_\tau\bar\nu_\tau$ channels.
The coefficients of $\delta_{\text{fitted}}$ for those two channels are given in Table.~\ref{tab::coefs}.
The comparison between the fitted functions and the full numerical results for 
the $b\bar b$ and $\nu_\tau\bar\nu_\tau$ channels are shown in the left and right panels of FIG.~\ref{fig::fittedz}, respectively. 
The relative deviations in the $b\bar b$ and $\nu_\tau\bar\nu_\tau$ channels are smaller than those in the $\tau^+\tau^-$ channel with the same input parameters. 
For the $Z\to c\bar c$ channel, even taking $\lambda^{1R}_{c\tau}=\sqrt{4\pi}$ and $M_{S_1}=1~{\rm TeV}$, the maximal deviation is only about $0.05\%$. 
Since this effect is much smaller than the corresponding deviations in the $\tau^+\tau^-$, $b\bar b$ and $\nu_\tau\bar\nu_\tau$ channels, we do not discuss the $Z\to c\bar c$ channel further.
Therefore, at the same level of measurement precision, the $\tau^+\tau^-$ channel provides a more sensitive probe of the scalar LQ $S_1$.

\bibliographystyle{JHEP}
\bibliography{refv2.bib}

\end{document}